\documentclass[twocolumn, reprint, aps, amsmath, amssymb, balancelastpage]{revtex4-2}
\pdfoutput=1
\usepackage{CJKutf8}

\usepackage{float}

\usepackage{graphicx}
\usepackage{dsfont}
\usepackage{amsmath}

\usepackage{array}

\usepackage{enumerate}
\usepackage{shuffle}

\usepackage{youngtab}
\usepackage{tensor}
\usepackage{braket}
\usepackage[normalem]{ulem}
\usepackage{slashed}
\usepackage[aligntableaux=center]{ytableau}
\usepackage[utf8]{inputenc}
\usepackage[
      colorlinks=true,
      linkcolor=blue,
      urlcolor=blue,
      filecolor=black,
      citecolor=red,
      ]{hyperref}
\usepackage{braket}
\usepackage{mathtools}
\usepackage{rotating}
\usepackage{tabularx}
\newcolumntype{Y}{>{\centering\arraybackslash}X}
\newcolumntype{C}[1]{>{\centering\arraybackslash}p{#1}}
\usepackage{todonotes}
\usepackage{xcolor}
\definecolor{LightCyan}{rgb}{0.7,1,1}
\definecolor{Gray}{gray}{0.9}
\usepackage{xspace}

\newcommand {\be} {\begin {equation}}
\newcommand {\ee} {\end {equation}}
\newcommand {\nn} {\nonumber}
\newcommand {\bes} {\begin {equation*}}
\newcommand {\ees} {\end {equation*}}

\newcommand{\es}[2] {\begin{equation} \label{#1} \begin{split} #2 \end{split} \end{equation}}

\newcommand{\cC}{{\mathcal C}}

\newcommand{\cG}{{\mathcal G}}
\newcommand{\cH}{{\mathcal H}}

\newcommand{\cL}{{\mathcal L}}

\newcommand{\cO}{{\mathcal O}}

\newcommand{\cQ}{{\mathcal Q}}

\newcommand{\cS}{{\mathcal S}}

\newcommand{\beq}{\begin{equation}}
\newcommand{\eeq}{\end{equation}}

\def\ie{\begin{equation}\begin{aligned}}
\def\fe{\end{aligned}\end{equation}}

\def\<{\langle}
\def\>{\rangle}

\def\beg{\begin{equation}\begin{gathered}}
\def\eeg{\end{gathered}\end{equation}}

\def\bea{\begin{equation}\begin{aligned}}
\def\eea{\end{aligned}\end{equation}}

\begin{document} 
\begin{CJK*}{UTF8}{gbsn}

\title{The AdS$_3\times $S$^3$ Virasoro-Shapiro amplitude with RR flux}

\author{Shai M.~Chester}     
\affiliation{
Blackett Laboratory, Imperial College, Prince Consort Road, London, SW7 2AZ, U.K.
		}
\author{De-liang Zhong (钟德亮)}
\affiliation{
Blackett Laboratory, Imperial College, Prince Consort Road, London, SW7 2AZ, U.K.
		}

\begin{abstract}
We compute the AdS Virasoro-Shapiro amplitude for scattering of dilatons in type IIB string theory with pure RR flux on $AdS_3\times S^3\times M_4$ for $M_4=T^4$ or $K3$, to all orders in $\alpha'$ in a small AdS curvature expansion. This is achieved by comparing the flat space limit of the dual D1D5 CFT correlator to an ansatz for the amplitude as a worldsheet integral in terms of single valued multiple polylogarithms. The first curvature correction is fully fixed in this way, and satisfies consistency checks in the high energy limit, and by comparison of the energy of massive string operators to a semiclassical expansion. Our result gives infinite predictions for CFT data in the planar limit at strong coupling, which can guide future integrability studies.
\end{abstract}

\maketitle
\end{CJK*}

\section{Introduction}

One of the original and simplest examples of the AdS/CFT correspondence relates the 2d D1D5 conformal field theory (CFT) to type IIB string theory on $AdS_3\times S^3\times M_4$, where $M_4$ can be either K3 or $T^4$ \cite{Maldacena:1997re}. The theories are related by the dictionary
\es{dict}{
\frac{R^2}{\ell_s^2}=g N\equiv\sqrt{\lambda}\,,
}
where $R$ is the AdS radius, $\ell_s^2=\alpha'$ is the string length, $N=\sqrt{Q_1Q_2}$ for $Q_1$ D1 and $Q_5$ D5 branes, and $g$ is the string coupling in six dimensions, which is proportional to both RR and NS-NS flux. Since 2d CFTs are more symmetric than higher dimensions, one might hope for a precise understanding of the duality, which indeed has been achieved recently in the tensionless small $\lambda$ limit of the $T^4$ case \cite{Eberhardt:2019ywk,Aharony:2024fid}, when there is no RR flux and minimal NS-NS flux. For small amounts of RR flux, integrability has been used to study the spectrum of the $T^4$ theory in the planar large $N$ limit \cite{Babichenko:2009dk,Hoare:2013lja,Borsato:2013qpa,Lloyd:2014bsa,Frolov:2023pjw,Hoare:2013pma,Borsato:2012ud,Borsato:2013hoa,OhlssonSax:2011ms,Borsato:2014exa,Borsato:2012ss,OhlssonSax:2012smh,Beccaria:2012kb,Cavaglia:2022xld,Ekhammar:2021pys,Cavaglia:2021eqr}. For a comprehensive list of references, see the reviews \cite{Sfondrini:2014via, Demulder:2023bux, Seibold:2024qkh}. However, for the most physically interesting case of large RR flux, i.e. large $\lambda$, where the bulk has a weakly curved gravity dual, little is known beyond the supergravity limit \footnote{A quantum spectral curve for the $T^4$ case was recently proposed in \cite{Cavaglia:2021eqr,Ekhammar:2021pys}, see \cite{Seibold:2024qkh} for a review, but is not yet at the point where CFT data can be computed the large $\lambda$ limit, unlike the better understood $AdS_5\times S^5$ case \cite{Gromov:2013pga}}. For instance, the scaling dimension of the simplest single trace operators dual to massive string states in the bulk, which are expected to scale as $\lambda^{1/4}$ in the planar large $\lambda$ limit, have not been determined.

Recently, a new method of computing scattering in AdS with large RR flux was developed originally for AdS$_5$/CFT$_4$ in \cite{Alday:2022uxp,Alday:2022xwz,Alday:2023jdk,Alday:2023mvu}. They considered the planar limit of scattering of gravitons in $AdS_5\times S^5$, which in the flat space limit is described by the Virasoro-Shapiro (VS) amplitude. They first applied a Borel transform to the block expansion of the dual CFT correlator, which generalizes the flat space limit of \cite{Penedones:2010ue} to include AdS curvature corrections. They then assume an ansatz of single valued multiple polylogarithms (SVMPLs) for the worldsheet integral of the AdS correlator. After comparing this ansatz to the Borel transformed block expansion, they were able to fix the first couple AdS curvature correction to all orders in $\alpha'$ \footnote{See \cite{Chester:2024esn} for similar results in $AdS_4\times \mathbb{CP}^3$ and the AdS Veneziano amplitude \cite{Alday:2024yax,Alday:2024ksp}.}, which was checked using previously known results from integrability \cite{Gromov:2013pga} and supersymmetric localization \cite{Chester:2020dja,Binder:2019jwn}.

Here, we apply this method to tree level scattering of dilatons in $AdS_3\times S^3\times M_4$, which is dual to the correlator the unique T-duality invariant half-BPS multiplet in the 2d CFT in the planar large $N$ limit but all orders in $1/\lambda$. The dilaton is not sensitive to $M_4$ at tree level, so this holographic correlator is the same for the K3 and $T^4$ CFTs, which both have small $\mathcal{N}=4$ superconformal symmetry \footnote{Note that the $T^4$ and K3 theories have 84 and 20 moduli, respectively. Each $\Delta=1$ half-BPS multiplet contains 4 marginal top components, so there are 21 and 5 such multiplets for each theory, respectively. We consider the half-BPS multiplet in each theory whose top components include the marginal operator that couples to the string coupling (as well as three other marginal operators related by $SO(4)$ automorphism), which is dual to the dilaton in the bulk. This is the unique half-BPS multiplet that does not care about the compact spacetime in the planar limit, and is invariant under the remnant of the duality symmetry (i.e. T-duality) that survives in the planar limit. The other half-BPS multiplets are related by these duality symmetries.}. The leading large $N$ and large $\lambda$ correction to this correlator, given by supergravity, was recently computed in \cite{Rastelli:2019gtj,Giusto:2018ovt} using the analytic bootstrap \footnote{See \cite{Galliani:2017jlg,Bombini:2017sge,Bombini:2019vnc,Giusto:2019pxc,Giusto:2020neo,Wen:2021lio,Behan:2024srh} for other supergravity results of holographic correlators in AdS3.}. The flat space limit (i.e. $\mathbb{R}^6\times M_4$) is also related to the VS amplitude, which for our kinematics takes the form
\es{flatVS}{
A^{(0)}(S,T)=-\frac{\hat\sigma_2\Gamma(-S)\Gamma(-T)\Gamma(-U)}{4\Gamma(S+1)\Gamma(T+1)\Gamma(U+1)}\,,
}
where we define the Mandelstam variables $S=-\frac{(p_1+p_2)^2}{4\sqrt{\lambda}}$, $T=-\frac{(p_1+p_3)^2}{4\sqrt{\lambda}}$, $U=-S-T$, and $\hat\sigma_2=S^2+T^2+U^2$. By applying a Borel transform to the holographic correlator and relating to an ansatz for the worldsheet integral in terms of SVMPLs, we are able to uniquely fix the first AdS curvature correction to this VS amplitude. Using this result, we are able to read off infinite CFT data in the large $N$ and large $\lambda$ limit. For instance, we compute the scaling dimension of the lowest Regge trajectory of massive string operators to subleading order, which we can check against previous partial results from a semiclassical expansion \cite{Beccaria:2012pm}. We also find that the high energy limit of our correlator is identical to the $AdS_5\times S^5$ \cite{Alday:2023pzu} and $AdS_4\times \mathbb{CP}^3$ \cite{Chester:2024esn} VS amplitudes.

The rest of this paper is organized as follows.  In Section \ref{sec:corr}, we discuss the correlator of half-BPS operators, including its block expansion, the Mellin transform, and previous results for the supergravity limit. In Section \ref{sec:flat}, we discuss the flat space limit of the holographic correlator, including AdS curvature corrections, and we fix the first couple orders of CFT data of massive string operators. In Section \ref{sec:worldsheet}, we write down our worldsheet ansatz for the first curvature correction, which we fix by comparison to the flat limit of the block expansion. In Section \ref{sec:check} we describe consistency checks from the high energy limit and the semiclassical analysis of the leading Regge trajectory. We end with a discussion of our results in Section~\ref{sec:discussion}. Additional details are given in the appendices and attached \texttt{Mathematica} file.

\section{Holographic correlator}\label{sec:corr}

We consider the four-point function of the superprimary $\cO(x;y,\bar y)$ of the $\frac12$-BPS multiplet discussed above, which is a $\Delta=1$ scalar in the $(\frac12,\frac12)$ of the $SU(2)\times SU(2)$ R-symmetry, and $y,\bar y$ are null polarization spinors. Superconformal symmetry constrains the correlator to \cite{Rastelli:2019gtj}
\es{4point}{
\langle\cO\cO\cO\cO\rangle=\frac{\cG_0+(1-z\alpha)(1-\bar z\bar\alpha)\cH(U,V)}{x_{12}^2x_{34}^2 y^{-1}_{12}y^{-1}_{34}\bar y^{-1}_{12}\bar y^{-1}_{34}}\,,
}
where we define the conformal cross ratios $\alpha=\frac{y_{13}y_{24}}{y_{12}y_{34}}$, $\bar\alpha=\frac{\bar y_{13}\bar y_{24}}{\bar y_{12}\bar y_{34}}$, $U=\frac{x_{12}^2x_{34}^2}{x_{13}^2x_{24}^2}$, $V=\frac{x_{23}^2x_{14}^2}{x_{13}^2x_{24}^2}$, the term $\cG_0(z,\bar z,\alpha,\bar\alpha)$ is protected, and the reduced correlator $\cH(U,V)$ is an R-symmetry singlet that contains all the unprotected dynamical information. The reduced correlator can be expanded in global conformal blocks as \cite{Aprile:2021mvq}
\es{blockExp}{
\cH(U,V)=\text{prot.}+\sum_{\Delta,\ell}C^2_{\tau\ell}U^{-1}g_{\tau+2,\ell}(U,V)\,,
}
where the first term refers to protected multiplets that will not matter for our purposes, while the second term includes the sum over all long multiplets whose superprimary has even spin $\ell$, twist $\tau=\Delta-\ell$, and OPE coefficient $C_{\tau,\ell}$. Here, $g_{\tau,\ell}(U,V)$ denotes a 2d global block
\es{2dblock}{
g_{\tau,\ell}(U,V)=\frac{k_{\tau+2\ell}(z)k_{\tau}(\bar z)+k_{\tau+2\ell}(\bar z)k_{\tau}( z)}{1+\delta_{\ell,0}}\,,
}
where $k_\beta(z)=z^{\beta/2}{}_2F_1(\beta/2,\beta/2,\beta;z)$.

The full correlator is an independent function of $z$ and $\bar z$, defined as $U=z\bar z$ and $V=(1-z)(1-\bar z)$, while $\cH(U,V)$  is a function of $U,V$, and thus symmetric in $z\leftrightarrow \bar z$. We can thus define the Mellin transform $M(s,t)$ of $\cH(U,V)$ as \cite{Rastelli:2019gtj} \footnote{A similar Mellin transform can be defined for the components of the full correlator that are symmetric in $z\leftrightarrow \bar z$, and can thus be written in terms of $U,V$ \cite{Rastelli:2019gtj}.}
\es{Mellin}{
\cH(U,V)&=\int\frac{dsdt}{(4\pi i)^2}U^{\frac s2}V^{\frac t2-1}M(s,t)\\
&\quad\times\Gamma\left(\frac{2-s}{2}\right)^2\Gamma\left(\frac{2-t}{2}\right)^2\Gamma\left(\frac{2-u}{2}\right)^2\,,
}
where $u=2-s-t$, and crossing symmetry swaps $s\leftrightarrow t\leftrightarrow u$. The Mellin transform of the block expansion in \eqref{blockExp} has poles in $s$ of the form
\es{blockM}{
M(s,t)\vert_{\text{$s$-poles}}=\text{prot.}+\hspace{-.05in}\sum_{\Delta,\ell} C^2_{\tau\ell} \hspace{-.05in}\sum_{m=0}^\infty \hspace{-.05in}  \frac{\cQ_{\tau+2,\ell,m}(t-2)}{s-\tau-2m}\,,
}
where we define the Mack polynomials $\cQ_{\tau,\ell,m}(x)$ in Appendix \ref{sec:blocks} \footnote{Note that our definition of Mack polynomials differs from the standard definition in \cite{Dey:2017fab}, due to the $U^{-1}$ factor in our blocks.}.

At large $N$ we can expand the correlator as $M=\frac{1}{N}M_\text{tree}+O(1/N^2)$, and the tree level term is constrained by the analytic bootstrap to take the form \cite{Rastelli:2019gtj}
\es{MlargeN}{
M_\text{tree}=M_\text{SG}+\hspace{-.05in}\sum_{a,b=0}^\infty \frac{\sigma_2^a\sigma_3^b}{\lambda^{\frac12+a+\frac32b}} \Big[\alpha_{a,b}^{{(0)}}+\frac{\alpha_{a,b}^{{(1)}}}{\sqrt{\lambda}}+\dots \Big],
}
where $\sigma_2=(s-\frac23)^2+(t-\frac23)^2+(u-\frac23)^2$ and $\sigma_3=(s-\frac23)(t-\frac23)(u-\frac23)$.
The leading term corresponds to an exchange Witten diagram given by supergravity, and was computed in \cite{Rastelli:2019gtj} to be \footnote{Their result was considered specifically for the K3 theory, and for tensor multiplets with an $SO(21)$ index, which is an approximate symmetry in the supergravity limit. We only consider a correlator of the specific $\frac{1}{2}$-BPS multiplet that does not care about the compact factor. This correlator was earlier computed by taking the light limit of a heavy-heavy-light-light correlator in \cite{Giusto:2018ovt}. }
\es{Msugra}{
M_\text{SG}(s,t)=-1/s-1/t-1/u\,,
}
where we chose the overall normalization to match \eqref{flatVS} in the flat space limit below.
The $1/\lambda$ corrections are polynomials in $s,t,u$ that correspond to contact Witten diagrams for higher derivative corrections to supergravity, and their coefficients $\alpha_{a,b}^{{(i)}}$ are not yet known. We organize these coefficients according to an expansion around the flat space limit, which we will discuss in the next section. We can also organize the expansion at finite AdS curvature but low energy, such that the first correction to supergravity comes from an $R^4$ term:
\es{R4}{
M_{R^4}=\frac{1}{\lambda^{\frac32}}\Big[\sigma_2\alpha_{1,0}^{(0)}+\alpha_{0,0}^{(2)}\Big]\,,
}
and so requires the second curvature correction.

\section{Flat space expansion}\label{sec:flat}

Consider the following Borel transform of the Mellin amplitude
\es{Borel}{
A(S,T)\hspace{-.025in}=\hspace{-.025in}{\lambda}^\frac12\hspace{-.05in}\int\hspace{-.05in}\frac{d\alpha}{2\pi i}\frac{e^\alpha}{\alpha^{3}}M\Big(\frac{2\sqrt{\lambda}S}{\alpha}+\frac23,\frac{2\sqrt{\lambda}T}{\alpha}+\frac23\Big),
}
where the shift by $2/3$ is so we can define $U=-S-T$ as is standard for flat space Mandelstam variables. The leading large $\lambda$ term of this Borel transform is just the flat space limit introduced in \cite{Penedones:2010ue}, for a correlator in 2d of external operators with $\Delta=1$ \footnote{Note that the Mellin amplitude of the reduced correlator is shifted by two powers of $s,t,u$ relative to the full correlator \cite{Rastelli:2019gtj}, which is why we have $\alpha^{-3}$ instead of $\alpha^{-1}$ in the integrand. The shifts of $2/3$ also disappears in the flat space limit.}. We define $A(S,T)$ as the AdS amplitude, which is expanded around flat space as 
\es{flatExp}{
A(S,T)&=A^{(0)}(S,T)+\frac{1}{\sqrt{\lambda}}A^{(1)}(S,T)+\dots\,,\\
A^{(k)}(S,T)&=\text{SG}^{(k)}+\sum_{a,b=0}^\infty  \frac{4^a8^b  \alpha_{a,b}^{(k)}  \hat \sigma_2^a\hat \sigma_3^b    }{\Gamma(2a+3b+3)} \,,\\
\text{SG}^{(0)}&=\frac{\hat\sigma_2}{4\hat\sigma_3}\,,\quad \text{SG}^{(1)}=\frac{\hat\sigma_2^2}{24\hat\sigma_3^2}\,,\quad\dots\,,
}
where we define $\hat\sigma_2=(S^2+T^2+U^2)$, $\hat\sigma_3=STU$, and $A^{(0)}(S,T)$ is the flat-space VS amplitude \eqref{flatVS}. We can now interpret the $\alpha_{a,b}^{{(k)}}$ coefficients in \eqref{MlargeN} as corresponding to the $k$th AdS curvature correction to the Wilson coefficients. From \eqref{flatVS}, we can read off the flat space terms, for instance
\es{alph0}{
\alpha^{(0)}_{a>0,0}=\frac{\Gamma(2a+3)}{8^{a}}\zeta(1+2a)\,,
}
while $\alpha^{(0)}_{0,0}=0$, as expected since the first higher derivative correction is $R^4$ whose flat space coefficient is $\alpha^{(0)}_{1,0}$. 

Our goal is to determine the first curvature correction $A^{(1)}(S,T)$, as well as the corresponding Wilson coefficients $\alpha^{(1)}_{a,b}$. The flat space limit is sensitive to high twist operators, which is why we only consider long operators. Following \cite{Alday:2022uxp}, we can expand the CFT data of these long operators as
\es{CFTdata}{
\tau(\delta,\ell)&=\tau_0(\delta,\ell)\lambda^{\frac14}+\tau_1(\delta,\ell)+\tau_2(\delta,\ell)\lambda^{-\frac14}+\dots\,,\\
C^2_{\tau,\ell}&=\frac{\pi^3(\delta_{\ell,0}+1)\lambda^{5/4}\tau^{-2}(\delta,\ell)}{4^{\ell+\tau(\delta,\ell)}\sin^2(\frac\pi2\tau(\delta,\ell))}f(\delta,\ell)\,,\\
f(\delta,\ell)&=f_0(\delta,\ell)+\frac{f_1(\delta,\ell)}{\lambda^{1/4}}+\frac{f_2(\delta,\ell)}{{\lambda}^{1/2}}+\dots\,,
}
where $\delta=1,2,\dots$ is the flat space string mass level, and the normalization of the OPE coefficients will be justified below. To fix the leading order coefficients, we need to take the Borel transform \eqref{Borel} of the block expansion \eqref{blockM} at large $\lambda$, and compare to the flat space VS amplitude \eqref{flatVS}. In Appendix \ref{sec:blocks}, we compute the poles in $S$ of the Borel transformed blocks in \eqref{blockM} and find
\es{blocksA}{
A_{\tau,\ell}(S,T)\vert_{S-\text{poles}}=\frac{4^{\ell+\tau}\tau^2\sin^2(\frac{\pi\tau}{2})}{(\delta_{\ell,0}+1)\lambda^{5/4}}\sum_{i=0}^\infty \frac{R^{(i)}_{\tau,\ell}(S,T)}{\lambda^{i/4}}\,,
}
where the first term is 
\es{Rfirst}{
R^{(0)}_{\tau,\ell}(S,T)=\frac{16T_\ell(1+\frac{2T}{S})}{\pi^3S^2\tilde\tau(\tilde\tau^2-4S)}\,,
}
where $\tilde\tau\equiv\tau/\lambda^{1/4}$, and $T_n(x)$ are Chebyshev polynomials, as expected for partial waves in 3d flat space. By comparing these poles against those of the flat space VS amplitude \eqref{flatVS}, we find that
\es{flatdata}{
\tau_0(\delta,\ell)=2\sqrt{\delta}\,,
}
and $f_{\delta,\ell}=0$ for $2\delta<\ell$. We give a large list of the explicit nonzero values in the attached \texttt{Mathematica} file. Note that only the leading Regge trajectory $\delta=\ell/2$ is unique \footnote{It would be interesting to compute the precise degeneracy of the higher Regge trajectories, as was done recently for the $AdS_5\times S^5$ case in \cite{Alday:2023flc}.}, so for higher Regge trajectories we can only compute the average of degenerate operators that we denote with angle brackets, e.g. $\langle f\rangle_{\ell/2-1,\ell}$. Curiously, we find that the leading Regge trajectory starts at $\ell=2$, unlike $AdS_5\times S^5$ or $AdS_4\times \mathbb{CP}^3$ where its starts at $\ell=0$ \footnote{For the pure NS-NS flux case, this fact can be explained by the explicit worldsheet construction in \cite{Ferreira:2017pgt}, where one sees that if one considers the class of states that correspond to the lowest Regge trajectory for spin $\ell$, then the $\ell\to0$ limit corresponds to a protected multiplet, and no longer a long multiplet. This is analogous to how in $\mathcal{N}=4$ super-Yang-Mills, the $\ell\to-2$ limit of the lowest Regge trajectory of long multiplets corresponds to a protected multiplet.}.

At the next order $1/\lambda^{1/4}$ we expect the AdS amplitude to vanish. We can impose this using the next term $R^{(1)}_{\tau,\ell}(S,T)$ given in \eqref{otherR}, from which we find that
\es{nextOrder}{
\tau_1(\delta,\ell)=-\ell-1\,,\quad \langle f_1\rangle_{\delta,\ell}=-\frac{(5+4\ell)\langle f_0\rangle_{\delta,\ell}}{4\sqrt{\delta}}\,.
}
The shift $\tau_1(\delta,\ell)=-\ell-1$ implies that the string states we consider are superdescendents with $\Delta$ shifted by one relative to the superprimary, as we will see from comparison to the semiclassical string state in the next section.

\section{Worldsheet integral}\label{sec:worldsheet}

For the next order $1/\sqrt{\lambda}$, we make an ansatz for the AdS amplitude $A^{(1)}(S,T)$ in terms of a worldsheet integral. Following \cite{Alday:2023jdk}, we assume that 
\es{ansatz1}{
A^{(k)}(S,T)=\int \frac{d^2z |z|^{-2S-2}}{|1-z|^{2T+2}}G^{(k)}(S,T,z)+\text{cross}\,,
}
where ``cross'' denotes the $S\leftrightarrow T$ and $S\leftrightarrow U$ terms. For instance, the leading VS amplitude is 
\es{anLead}{
G^{(0)}(S,T,z)=\frac{\hat\sigma_2}{12U^2}\,.
}
The ansatz for the integrand $G^{(1)}(S,T,z)$ is written in terms of SVMPLs as
\es{ansatz2}{
G^{(1)}=\sum_{i}p_i^s(S,T)\cL^s_i(z)+\sum_{j}p^a_j(S,T)\cL^a_j(z)\,,
}
where $p^{s/a}(S,T)$ are symmetric/antisymmetric degree two polynomials, while $\cL^{s/a}(z)$ are symmetric/antisymmetric combinations of weight three SVMPLs
\es{svmpls}{
\cL^s&=\{\cL^s_{000},\cL_{001}^s,\cL_{010}^s,\zeta(3)\}\,,\\
\cL^a&=\{\cL^a_{000},\cL_{001}^a,\cL_{010}^a\}\,,\\
}
whose expressions we review in Appendix \ref{sec:svmpls}. The relative weight between the polynomials and SVMPLs is fixed by the expected transcedentality of coefficients in the low energy expansion \eqref{MlargeN} \footnote{Note that our functions of $S,T$ are two degrees higher than that of \cite{Alday:2023jdk}, because the reduced correlator in 4d $\mathcal{N}=4$ SYM differs by two degrees in $s,t$ than the reduced correlator in our case.}. The weight three follows from the argument in \cite{Alday:2023jdk} about how expanding the worldsheet metric into AdS around flat space induces extra gravitons, with each extra graviton increasing the weight by three. 

In sum, we have 11 coefficients in our worldsheet integrand ansatz $G^{(1)}(S,T,z)$, of which two are ambiguities that integrates to zero in $A^{(1)}(S,T)$, as we show in Appendix \ref{sec:svmpls}. We can fix these coefficients by taking poles in $S=1,2,\dots$, performing the $z$ integrals as detailed in Appendix \ref{sec:svmpls}, and comparing to the poles in the block expansion \eqref{blocksA}, where we use the next term $R^{(2)}_{\tau,\ell}(S,T)$ as given in the attached \texttt{Mathematica} file. This fixes 8 of the parameters. One more parameter is fixed by taking the $S=0$ pole and comparing to the known result for the supergravity term $\text{SG}^{(1)}(S,T)$ in \eqref{flatExp}. The last two parameters are ambiguities that we simply set to zero. The result is
\es{answer}{
&p^s=\Big\{\frac{1}{192} \left(9 \left(S^2+T^2\right)+52 S T\right)\,,\,-\frac{13 S
   T}{48},\\
   &\qquad-\frac{1}{96} \left(12 S^2+25 S T+12 T^2\right)\,,\,
   \left(S^2+T^2\right)\Big\}\,,\\
   &p^a=(S^2-T^2)\left\{{3}/{64} \,,\,{5}/{24}
  \,,\,{5}/{48} \right\}\,.\\
}
Our solution also fixes the $\tau_2(\delta,\ell)$ and $f_2(\delta,\ell)$ coefficients in the CFT data \eqref{CFTdata}, which we show explicitly in the attached \texttt{Mathematica} file. For instance, for the unique lowest Regge trajectory we find
\es{lowReg}{
\tau_2(\ell/2\,,\ell)=\frac{3 \ell^2-2 \ell+2}{4 \sqrt{2} \sqrt{\ell}}\,.
}
We can then take the low energy expansion of $A^{(1)}(S,T)$ as shown in Appendix \ref{sec:svmpls} to get
\es{lowA1}{
A^{(1)}(S,T)=&\frac{\hat\sigma_2^2}{24\hat\sigma_3^2}
-\frac{57}{4}
   \zeta (5) \hat\sigma_3
   -\frac{31}{3} \hat\sigma_2^2 \zeta
   (3)^2
  \\
& -\frac{1089}{16} \zeta (7) \hat\sigma_3 \hat\sigma_2+\dots\,,
}
which we can use to fix the coefficients $\alpha_{a,b}^{(1)}$ in \eqref{flatExp} and \eqref{MlargeN} to
\es{lowEnergy}{
\alpha^{(1)}_{0,1}&=-\frac{855}{4}\zeta(5)\,,\quad  \alpha^{(1)}_{2,0}=-465\zeta(3)^2\,,\\
\alpha^{(1)}_{1,1}&=-\frac{343035}{32}\zeta(7)
 \,,\quad\dots\,,
}
and it is straightforward to go to higher order. Note that we get a nonzero correction to the $D^4R^4$ term $\zeta(5)\hat\sigma_3$, unlike the $AdS_5\times S^5$ case where this correction vanishes \cite{Alday:2022uxp} \footnote{The flat space $D^4R^4$ is of course nonzero in both cases; only the AdS curvature correction differs. The non-vanishing of the $D^4R^4$ correction for $AdS_5\times S^5$ was shown both by the AdS VS calculation in \cite{Alday:2022uxp}, and also by the more rigorous calculation combining analytic bootstrap with supersymmetric localization in \cite{Chester:2020dja}}.

\section{Consistency checks}\label{sec:check}

We have two checks on our solution. The first comes from the high energy limit, where we take $|S|,|T|\gg1$ with $S/T$ fixed. As shown in \cite{Alday:2023pzu}, we do this by taking the saddle point $z_0=\frac{S}{S+T}$ in the worldsheet integral \eqref{ansatz1}. This gives
\es{high}{
A^{(1)}_\text{HE}(S,T)=A^{(0)}_\text{HE}(S,T) W_3(z_0)\,,
} 
where the flat space VS with our normalization is
\es{high0}{
A^{(0)}_\text{HE}(S,T) =\frac{\hat\sigma_2}{4U^2}e^{-2S\log|S|-2T\log|T|-2U\log|U|}\,,
}
and the next correction has
\es{W3}{
W_3(z_0)=&\cL_{000}(z_0)-\cL_{001}(z_0)-\frac{1}{z_0}\cL_{010}(z_0)\\
&-\frac{(z_0-1)^2}{z_0^2}\cL_{011}(z_0)+\frac{z_0-1}{z_0^2}\cL_{101}(z_0)\\
&+\frac{(z_0-1)^2}{z_0^2}\cL_{111}(z_0)+2\zeta(3)\,.
} 
This is in fact the exact same answer as in the $AdS_5\times S^5$ \cite{Alday:2023pzu} and $AdS_4\times \mathbb{CP}^3$ \cite{Chester:2024esn} cases, and so in particular it satisfies the exponentiation requirement discussed in \cite{Alday:2023pzu}.

The second check comes from comparing our result for the scaling dimension of the first Regge trajectory from the prediction from the semiclassical analysis in \cite{Forini:2012bb,Beccaria:2012pm} for short strings in $AdS_3\times S_3\times M_4$ with pure RR flux \footnote{The original references \cite{Forini:2012bb, Beccaria:2012pm} only considered $M_4=S^3\times S^1$ and $M_4=T^4$, but we show that the result does not care about the compact spacetime, and so is the same for $M_4=K3$.}. As we review in Appendix \ref{sec:semi}, one can compute the scaling dimension of massive stringy states by computing the worldsheet action semiclassically where we assume the spin $\ell$ and the $SU(2)\times SU(2)$ isospins $(\frac J2,\frac J2)$ are large, and then extrapolating to finite $\ell$ and $J$. Similar extrapolations in $AdS_5\times S^5$ matched later rigorous calculations from integrability \cite{Roiban:2011fe}. In our case, we get
\es{semiAns}{
E=\lambda^{\frac14}\sqrt{2\ell}+\lambda^{-\frac14}\Big[\frac{3\ell^{\frac32}}{4\sqrt{2}}+\frac{J^2}{2\sqrt{2\ell}}+\cC \sqrt{2\ell} \Big]+\dots\,,
}
where $\cC$ is a 1-loop coefficient that was not yet computed \footnote{The extrapolation can be subtle. For instance, in the $AdS_4\times \mathbb{CP}^3$ case, the $\frac{J^2}{2\sqrt{2\ell}}$ term receives quantum corrections and becomes $\frac{(J+1/2)^2}{2\sqrt{2\ell}}$, as discussed in \cite{Gromov:2014eha}. In the $AdS_5\times S^5$ case though, and apparently also in our case, there are no such additional shifts.}. This matches our results \eqref{flatdata} and \eqref{lowReg} provided that we set $J=1$, which implies that the stringy state is a descendent of the $J=0$ superprimary with $\Delta$ shifted by one, as shown in Appendix \ref{sec:semi}, and as we also observed in \eqref{nextOrder}. Our result also gives a prediction for the 1-loop coefficient
\es{C}{
\cC=-{1}/{4}\,,
}
which curiously is the same value computed semiclassically for $AdS_5\times S^5$ in \cite{Roiban:2011fe}. Note the classical string solutions as a function of $J$ and $\ell$ are the same for both $AdS_5\times S^5$ and $AdS_3\times S^3$, as we discuss in the Supplementary material, but it is not obvious why the 1-loop correction should be the same. One-loop computations of the fluctuation frequencies around the semiclassical folded solution for  $M_4 = T^4$ have been carried out in \cite{Beccaria:2012kb, Beccaria:2012pm}. However, due to unresolved regularization mismatches, a definitive one-loop result could not be established.

\section{Discussion}\label{sec:discussion}

In this work we computed the first AdS curvature correction to the Virasoro-Shapiro amplitude on $AdS_3\times S^3\times M_4$ for $M_4=K3 \,,T^4$ and pure RR flux. Our result was completely fixed by assuming a worldsheet ansatz analogous to that of the $AdS_5\times S^5$ case \cite{Alday:2023jdk}, and we have two consistency checks by considering the high energy limit, and by comparison to a semiclassical calculation of the leading Regge trajectory in \cite{Beccaria:2012pm}. Our result fixes the 1-loop term in the semiclassical calculation, and gives infinite predictions for other CFT data of massive stringy operators that can be used to guide future integrability studies.

To compute the next curvature correction, we will likely need input from integrability, as was the case for  $AdS_5\times S^5$ in \cite{Alday:2023mvu} \footnote{In particular, we would need to know the scaling dimension of the lowest Regge trajectory at the order $1/\lambda^{3/4}$.}. This second curvature correction is necessary if we want to fix the first higher derivative correction $R^4$ at finite AdS curvature, as shown in \eqref{R4}. 

Our calculation opens to the door to many generalizations. We can consider correlators of higher KK modes, as was done in the $AdS_5\times S^5$ case in \cite{Fardelli:2023fyq}, as the leading supergravity terms were already computed in \cite{Rastelli:2019gtj}. We can also consider the case of mixed NS-NS and RR flux, where we can use the fact that the pure NS-NS flux correlator can in principle be computed independently from the worldsheet \footnote{The recent paper \cite{Alday:2024rjs} took a first step toward computing the pure NS-NS correlator, where they considered the correlator of four tachyons in bosonic string theory on $AdS_3$, and in the momentum basis (called the ``m''-basis \cite{Cardona:2010qf}). The next step would be to extend this calculation to the correlator of the lowest half-BPS multiplet in superstring theory on $AdS_3\times S^3\times T^4$, and to the position basis (called the ``x''-basis), and then take the Mellin transform \eqref{Mellin} and then the Borel transform \eqref{Borel}, which would give the pure NS-NS amplitude in the basis used in this and previous papers for the AdS Virasoro-Shapiro amplitude in various dimensions.}. We can also generalize to $M_4=S^3\times S^1$, which has a different superconformal group: the large $\mathcal{N}=4$ algebra. Even the supergravity term has not yet been computed in this case, as the relevant Ward identities have not yet been derived \footnote{The derivation of the Ward identities in the small $\mathcal{N}=4$ case in \cite{Rastelli:2019gtj} was done using superspace and also using the 2d chiral algebra limit, none of which are currently known for the large $\mathcal{N}=4$ algebra.}. Finally, we could consider correlator of other four half-BPS multiplets that are sensitive to the compact spacetime for the $T^4$ case (and transform as a vector under the $SO(4)$ subgroup of T-duality that becomes an approximate symmetry at tree level), which at tree level should still be described by the Virasoro-Shapiro amplitude since $T^4$ is flat, unlike the $K3$ case where we expect a more complicated correlator \cite{Lin:2015dsa}.

\section*{Acknowledgments}

We thank Ofer Aharony, Xi Yin, Fernando Alday, Tobias Hansen, Lorenz Eberhardt, Arkady Tseytlin, and Erez Urbach for useful conversations, and Ofer Aharony for reviewing the manuscript. SMC and DlZ are supported by the UK Engineering and Physical Sciences Research council grant number EP/Z000106/1, and the Royal Society under the grant URF\textbackslash R1\textbackslash 221310. 

\bibliography{AdS3.bib}

\newpage
\onecolumngrid
\appendix
\section*{Supplemental Materials}
\section{Flat space expansion of blocks}
\label{sec:blocks}

The Mellin transform \eqref{Mellin} of the expression $U^{-1}g_{\tau+2,\ell}(U,V)$ that appears in the position space block expansion \eqref{blockExp}has poles in $s$ given by
\es{melBlock}{
\sum_{m=0}^\infty\frac{\cQ_{\tau+2,\ell,m}(t-2)}{s-\tau-2m}\,,
}
where $\cQ_{\tau,\ell,m}(t)$ is defined as
\es{mack}{
\cQ_{\tau,\ell,m}(t)=K(2,\tau,\ell,m,2)Q_{\ell,m}^{\tau,d=2}(t)\,,
}
the polynomial $Q_{\ell,m}^{\tau,d}$ is defined in Appendix A of \cite{Alday:2022uxp} (and earlier in \cite{Dey:2017fab,Mack:2009mi}), and the coefficient for a correlator of external dimension $\Delta$ and a block of twist $\tau$ and spin $\ell$ in dimension $d$ with descendent $m$ is
\es{K}{
K(\Delta,\tau, \ell, m,d)  =  \frac{-2^{1-\ell}   \Gamma (\ell+\frac{\tau}{2} )^{-4}(\ell+\tau -1)_\ell \Gamma (2 \ell+\tau )}{m!\Gamma (\Delta -\frac{\tau }{2}-m)^2 \left(\ell+\tau-\frac{d}{2} +1\right)_m}.
}
Note that even though $\Delta_\cO=1$, we set $\Delta=2$ in \eqref{mack} in order to match the position space expression $U^{-1}g_{\tau+2,\ell}(U,V)$. We next take the Borel transform \eqref{Borel} of \eqref{melBlock} to get
\es{BorelBlock}{
A_{\tau,\ell}(S,T)\vert_{S-\text{poles}}={\lambda}^\frac12\hspace{-.025in}\int\hspace{-.025in}\frac{d\alpha}{2\pi i}\frac{e^\alpha}{\alpha^{3}}\sum_{m=0}^\infty\frac{\cQ_{\tau+2,\ell,m}(\frac{2\sqrt{\lambda}T}{\alpha}-\frac43)}{\frac{2\sqrt{\lambda}S}{\alpha}+\frac23-\tau-2m}\,.
}
For each $m$ we do the integral by picking the pole
\es{alphPole}{
\alpha_*=\frac{2\sqrt{\lambda}S}{\tau+2m-\frac23}\,,
}
which is the only contribution to the $\alpha$ integral that is relevant to the $S$ poles that we are ultimately interested in. We now want to sum over $m$ for large $\tau$. The Mack polynomial in this regime has its maximum at $m\sim \tau^2$, so we set $m=x\tau^2$ and replace the sum by an integral to get
\es{BorelBlock2}{
A_{\tau,\ell}(S,T)\vert_{S-\text{poles}}=-\frac{\tau^2}{2S}\hspace{-.025in}\int_0^\infty\hspace{-.025in}{dx}\frac{e^{\alpha_*}}{\alpha_*}{\cQ_{\tau+2,\ell,m}(\frac{2\sqrt{\lambda}T}{\alpha_*}-\frac43)} \,.
}
We now expand this at large $\lambda$, keeping in mind that $\tau\sim\lambda^{1/4}$, and do the $x$ integrals order by order to get the expansion in \eqref{blocksA}, where the first term was given in \eqref{Rfirst}, the second is
\es{otherR}{
R_{\tau,\ell}^{(1)}&=-\frac{8 \left(4 (2 \ell+3) S+(2 \ell+1)
   \tilde\tau^2\right)
   T_\ell\left(1+\frac{2 T}{S}\right)}{\pi ^3
   S^2 \tilde\tau^2 \left(\tilde\tau^2-4 S\right)^2},\\
}
and the third is much messier so we relegate it to the attached \texttt{Mathematica} file.

\section{Worldsheet calculations}
\label{sec:svmpls}

We make use of symmetrized/antisymmetrized weight three SVMPLs, which are defined as 
\es{symAS}{
\cL_w^s(z)&=\cL_w(z)+\cL_w(1-z)+\cL_w(\bar z)+\cL_w(1-\bar z)\,,\\
\cL_w^a(z)&=\cL_w(z)-\cL_w(1-z)+\cL_w(\bar z)-\cL_w(1-\bar z)\,,\\
}
while the explicit expression for weight three SVMPLs can be found in the Appendix of \cite{Alday:2023jdk}, and they are written in terms of products of Polylogarithms. 

In the ansatz \eqref{ansatz2} we use for the integrand of the worldsheet integral \eqref{ansatz1}, two linear combinations of the polynomials integrate to zero, and thus do not affect any physical observables. These ambiguities $p_{1,2}$ take the form
\es{ambs}{
p_1^s&=\big\{0,0,0,{ST}\big\}\,, \quad p_1^a=\{0,0,0\}\,,\\
p_2^s&=\big\{S^2+4ST+T^2,\quad-4(S^2+ST+T^2),\quad 4S^2-2ST+4T^2,\quad-32(S^2+T^2)\big\},\\
p_2^a&=(S^2-T^2)\{1,4,0\}\,.
}
In practice, after solving for nine of the eleven parameters, we can simply set the remaining parameters to zero.

To compute the poles in $S$ of the worldsheet ansatz, we must do the $z$ integrals only keeping the polar terms. We do this for pole $S=\delta$ following \cite{Alday:2023jdk} by changing to polar coordinates $z=\rho e^{i\theta}$, expanding the ansatz to order $2\delta$ in $\rho$, performing the trivial $\theta$ integrals, and obtaining the poles from integrals of the form
\es{rho}{
\int_0^{\rho_0} d\rho {\rho^{-2S+2\delta-1}\log^p(\rho^2)}=-\frac{p!}{2(S-\delta)^{p+1}}+\text{nonpolar}\,,
}
where the upper bound $\rho_0$ does not affect the polar terms. In practice, our weight three ansatz gives degree four poles and smaller, which is the same degree poles that appear in the block expansion.

For the low energy expansion, we can obtain the polar terms by the previous method for $\delta=0$. For the non-polar terms, we instead use the formula \cite{Alday:2023jdk}
\es{lowEng}{
&\int d^2z|z|^{-2S-2}|1-z|^{-2T-2}\cL_w(z)=\text{polar}+\sum_{p,q=0}^\infty(-S)^p(-T)^q\sum_{W\in 0^p \shuffle 1^q \shuffle w} \!\!\!\!\!\!\!\!(\cL_{0W}(1)-\cL_{1W}(1))\,,
}
where SVMPLs at $z=1$ can be evaluated using \texttt{PolyLogTools} \cite{Duhr:2019tlz}.

\section{Semiclassical analysis}
\label{sec:semi}

For the pure RR case, the semiclassical analysis of the $AdS_3 \times S^3 \times M_4$ background is similar to the well-studied case of $AdS_5 \times S^5$. Specifically, the metric of the $AdS_3 \times S^3 \times M^4$ background is given by:
\beq \label{eqn-metrics}
\begin{aligned}
ds^2 &=  ds_{AdS_3}^2 + ds_{S^3}^2 +  ds_{M_4}^2, \\ \nn
ds_{AdS_3}^2&= -\cosh^2{\rho} dt^2+d\rho^2+\sinh^2{\rho} d\phi^2, \\
ds_{S^3}^2&= d \beta_1^2+\cos^2{\beta_1}(d\beta_2^2+\cos^2{\beta_2}d\beta_3^2)\, ,    
\end{aligned}
\eeq
and the Polyakov action is given by 
\beq 
S=\frac{\sqrt{\lambda}}{4\pi}\int d\tau d\sigma \Big[\sqrt{-h}h^{ab}G_{mn}\partial_a X^m \partial_b X^n\Big]\,,
\eeq
where $h^{ab}=\text{diag}(-1,1)$. Here, the space-time indices $m, n$ run from $0$ to $5$, where the directions $0, 1, 2$ label $AdS_3$ while $3, 4, 5$ label the $S^3$. As the classical solutions we are considering have no dynamics on $M^4$ we will ignore those internal directions.

\paragraph{Folded String Solution.}

The classical string solution is characterized by the $AdS_3$ spin $\mathcal{S}$ and the two angular momenta $J_1, J_2$ on the $S^3$. The specific classical solution relevant to our analysis is the folded string solution, which has quantum numbers $(\cS, J_1, J_2) = (\cS, J, 0)$. This solution is identical to the folded string solution on $AdS_5 \times S^5$ \cite{Frolov:2002av}, as there the classical motion is taken in the $AdS_3 \times S^3$ subspace. 

One can first solve the energy of the classical string in the semiclassical regime $\cS,J = \mathcal{O}(\sqrt{\lambda}) $ and then extrapolate the dispersion relation to small $\cS,J$. The result is
\beq
E = \sqrt{2 \cS \sqrt{\lambda}} \left(1 +\frac{\frac{J^2}{4 \cS}+\frac{3 \cS}{8} + \mathcal{C}}{\sqrt{\lambda}} +\mathcal{O}\left(\frac{1}{\lambda}\right) \right)\, ,
\eeq
where $\mathcal{C}$ is a constant that could be computed from the 1-loop semiclassical analysis. 

\paragraph{Long Multiplet Fitting.}

The reason we believe that the folded string solution is the appropriate candidate is because it fits into the long multiplet of the superconformal primary that is an R-symmetry singlet.

Recall that the field theory dual to the $AdS_3 \times S^3 \times M^4$ background has small $\mathcal{N}=4$ superconformal symmetry. We consider the global subgroup of this symmetry, where all states are labeled by two $SU(2)$ R-symmetry quantum numbers $(j, \bar{j})$, and the two-dimensional scaling dimensions $(h, \bar{h})$, defined as:
\begin{equation}
h, \bar{h} \equiv \frac{\Delta \pm \ell}{2},
\end{equation}
where $\ell \in \mathbb{Z}/2$ is the Lorentz spin. We denote these states as:
\begin{equation}
[j, \bar{j}]_{h, \bar{h}}.
\end{equation}
The supercharges relevant to our analysis are discussed in \cite{Lee:2019uen}, and take the form:
\begin{equation}
G^{i \pm}_{-\frac{1}{2}} \in \left[\pm \frac{1}{2}, 0\right]_{\frac{1}{2}, 0}, \qquad \overline{G}^{i \pm}_{-\frac{1}{2}} \in \left[0, \pm \frac{1}{2}\right]_{0, \frac{1}{2}},
\end{equation}
where $i=1,2$ and the notation $\in$ denotes that $G^{\pm}$ transforms in the fundamental representation of $SU(2)$.

We are considering a superprimary that is an R-symmetry singlet, let us denote its quantum number by $[0,0]_{h_0,\bar{h}_0}$. To get bosonic descendants, we can act with an even numbers of the supercharges on the superprimary.

Our goal is to identify a suitable semiclassical state that, upon extrapolating its quantum numbers to small values, fits into the long multiplet. To achieve this, we first map the $SO(4)_R$ angular momenta $(J_1, J_2)$ introduced earlier to the $SU(2) \times SU(2)$ quantum numbers $(j, \bar{j})$, as follows:
\begin{equation}
j = \frac{J_1 + J_2}{2}, \quad \bar{j} = \frac{J_1 - J_2}{2}.
\end{equation}
Using this dictionary, the folded string state with Lorentz spin $\cS$ and $J= 1$ can be identified as
\begin{equation}
   \mathtt{Folded\ String} \equiv \left[\frac{1}{2}, \frac{1}{2}\right]_{\frac{\Delta + \cS}{2}, \frac{\Delta - \cS}{2}}\, ,
\end{equation}
which can fit into the supermultiplet of a superprimary with the form:
\begin{equation}
\left[0, 0\right]_{\frac{\Delta + \ell}{2}, \frac{\Delta - \ell}{2} },
\end{equation}
by acting on it with either one $G$ or three $G$s, and one $\overline{G}$ or three $\overline{G}$s, raising the R-charge by $[1/2,1/2]$. Explicitly, we have:
\begin{itemize}
   \item One $G$ and one $\overline{G}$:  
      \begin{equation}
      G^{i,+}_{-\frac{1}{2}} \overline{G}^{i,+}_{-\frac{1}{2}} \left[0,0 \right]_{\frac{\Delta + \ell}{2}, \frac{\Delta - \ell}{2} } = \left[\frac{1}{2}, \frac{1}{2}\right]_{\frac{\Delta + \ell+1}{2}, \frac{\Delta - \ell+1}{2}}.
      \end{equation}
      Here the choice of $i$ is arbitrary. To identify the state on the RHS with the folded string, we see that the spin of the superconformal primary is still $\ell$ but the conformal dimension is shifted by one, which agrees with \eqref{nextOrder}. So we should identify $\cS=\ell$, which gives \eqref{semiAns}.
   \item Three $G$s and one $\overline{G}$:
      \begin{equation}
         G^{1,+}_{-\frac{1}{2}} G^{1,-}_{-\frac{1}{2}} G^{2,+}_{-\frac{1}{2}} \overline{G}^{1,+}_{-\frac{1}{2}} \left[0,0 \right]_{\frac{\Delta + \ell}{2}, \frac{\Delta - \ell}{2} } = \left[\frac{1}{2}, \frac{1}{2}\right]_{\frac{\Delta + \ell+3}{2}, \frac{\Delta - \ell+1}{2}}.
      \end{equation}
      Here, we provide one option among the possible choices of selecting three $G$s and one $\overline{G}$. To identify the state on the RHS with the folded string, we note that the spin of the state is shifted relative to the superconformal primary as $\mathcal{S} = \ell + 1$, and the conformal dimension is increased by two. These states would thus have odd spin, which excludes the case we consider.
   \item Three $\overline{G}$s and one $G$: this case can be analyzed similarly to the previous case by exchanging $G$ and $\overline{G}$. 
   \item Three $G$s and three $\overline{G}$s: 
   \begin{equation}
      G^{1,+}_{-\frac{1}{2}} G^{1,-}_{-\frac{1}{2}} G^{2,+}_{-\frac{1}{2}} \overline{G}^{1,+}_{-\frac{1}{2}} \overline{G}^{1,-}_{-\frac{1}{2}} \overline{G}^{2,+}_{-\frac{1}{2}} \left[0,0 \right]_{\frac{\Delta + \ell}{2}, \frac{\Delta - \ell}{2} } = \left[\frac{1}{2}, \frac{1}{2}\right]_{\frac{\Delta + \ell+3}{2}, \frac{\Delta - \ell+3}{2}}.
   \end{equation}
   Here, we present one possible choice among several. To identify the state on the RHS with the folded string, we note that the spin of the superconformal primary remains the same, $\ell = \mathcal{S}$, while the conformal dimension is shifted by three. However, in our analysis so far, we do not observe this shift by three. Therefore, we will exclude this state from our consideration.\end{itemize}

\end{document}